\documentclass[aps,twocolumn,superscriptaddress,showpacs,floatfix]{revtex4-1}
\usepackage{amsfonts,amssymb,latexsym,xspace,epsfig,graphicx,color}
\usepackage{amsmath,enumerate,stmaryrd,xy,stackrel,ulem,multirow}
\usepackage[colorlinks=true,citecolor=blue,linkcolor=blue,urlcolor=blue]{hyperref}

\usepackage[usenames, svgnames]{xcolor}
\definecolor{OliveGreen}{rgb}{0,0.5,0}

\newcommand{\fref}[1]{Fig.~\ref{#1}}

\begin{document}
\title{
Coherent phase slips in coupled matter-wave circuits
}

\author{A. P\'erez-Obiol} 
\thanks{Both authors contributed equally to this work}
\affiliation{Barcelona Supercomputing Center, 08034 Barcelona, Spain}
\email{axel.perezobiol@bsc.es}
\author{J. Polo} 
\thanks{Both authors contributed equally to this work}
\affiliation{Quantum Research Centre, Technology Innovation Institute, Abu Dhabi, UAE}
\email{juan.polo@tii.ae}
\author{L. Amico}
\thanks{On leave from Dipartimento di Fisica e Astronomia 'Ettore Majorana', Universit\`a di Catania, Italy}
\affiliation{Quantum Research Centre, Technology Innovation Institute, Abu Dhabi, UAE}
\affiliation{INFN-Sezione di Catania, Via S. Sofia 64, 95127 Catania, Italy}
\affiliation{Centre for Quantum Technologies, National University of Singapore, 3 Science Drive 2, Singapore 117543, Singapore}
\affiliation{LANEF `Chaire d’excellence’, Université Grenoble-Alpes \& CNRS, F-38000 Grenoble, France}

\date{\today}

\begin{abstract}
Quantum Phase slips are  dual process of particle tunneling in coherent networks.  Besides to be of central interest for condensed matter physics, quantum phase slips are resources that are sought to be manipulated in  quantum circuits.  
Here, we devise a specific matter-wave circuit enlightening quantum phase slips. 
Specifically, we investigate the  quantum many body dynamics of  two  side-by-side ring-shaped neutral bosonic systems coupled through a weak link. By imparting a suitable magnetic flux, persistent currents flow in each ring with given winding numbers.  We demonstrate that coherent phase slips occur as winding number transfer among the two rings, with the populations in each ring remaining nearly constant. Such a phenomenon occurs as a result of a specific entanglement of circulating states, that, as such cannot be captured by a mean field treatment of the system. Our work can be relevant for the observation of quantum phase slips in cold atoms  experiments and their manipulation in  matter-wave circuits. To make contact with the field, we show that the phenomenon has clear signatures in the momentum distribution of the system providing the time of flight image of the condensate.

\end{abstract}

\maketitle

\paragraph*{Introduction}

Phase slips  are jumps of the phase of the wave function. In coherent systems as superconducting  and cold atoms networks, they   occur because of the suppression of the amplitude of the superconducting/superfluid order  parameter making the phase unrestricted and able to jump by a discrete amount (in multiples of $2\pi$) \cite{arutyunov2008superconductivity,d2017quantum}.  When such suppression  is caused by thermal fluctuations, a thermal phase slip occurs. Quantum Phase Slips (QPS) instead, are induced by  quantum fluctuations. A way to engineer QPS in mesoscopic physics is through  Josephson junctions, in which  such events correspond to tunneling of the phase of the order parameter \cite{arutyunov2008superconductivity,rastelli2013quantum}. Intriguingly, relying on a phase-charge duality,  it was argued that such tunneling events occur in proximity  of  the Coulomb blocked regime \cite{mooij2006superconducting,hriscu2011coulomb}. 

In cold atoms settings, QPS have been investigated in different settings with enhanced control and flexibility of the physical conditions \cite{polkovnikov2002nonequilibrium,polkovnikov2005decay,khlebnikov2005quantum,danshita2013universal,roscilde2016quantum}. Atomtronic circuits, in particular, define coherent networks  to study mesoscopic effects and  quantum transport of  ultracold atoms \cite{amico_focus_2017,amico2021roadmap,amico2021atomtronic,brantut2012conduction,burchianti2018connecting}. In the interesting configuration of bosonic condensates in a toroidal geometry, thermal and  QPS have been suggested as the main responsible for the dynamics of the atoms' persistent flow  through the nucleation of vortex  states  and  associated  phonon  emission, or through the formation of dark  solitons  activated  thermally  or by the stirring protocol \cite{wright2013driving,ramanathan2011superflow,eckel2014hysteresis,yakimenko2015vortices,polo2018damping,polo2019oscillations,perez2020bose,perez2020current}. 

Although over the past three decades several phenomena in superconducting networks have been related to  QPS' formation \cite{lau2001quantum,bezryadin2000quantum,bollinger2008determination,altomare2006evidence,masluk2012microwave,manucharyan2009fluxonium,weissl2015kerr,pop2010measurement}, a convincing  experimental evidence  of their occurrence in solid state physics has been obtained  only recently \cite{astafiev2012coherent}. In cold atoms instead, QPS, as a coherent transfer of vortices or flux, have not been observed yet. 

Besides their interest in fundamental physics, QPS can be a {\it resource} for quantum technology. Josephson junctions based quantum devices harnessing QPS have been carried out  \cite{mooij2005phase,belkin2015formation,pop2010measurement}. For atomtronic ring circuits interrupted by weak links,  defining the atomic counterparts of SQUID devices, QPS localized at the weak link play a crucial role for creating the superposition of the current states that is expected to be especially important for quantum sensing \cite{cominotti2014optimal,aghamalyan_atomtronic_2016,amico_superfluid_2015,ryu2013experimental,polo2019oscillations,polo2020quantum}.
\begin{figure}[tbh]
  \centering
  \includegraphics[width = 1\linewidth]{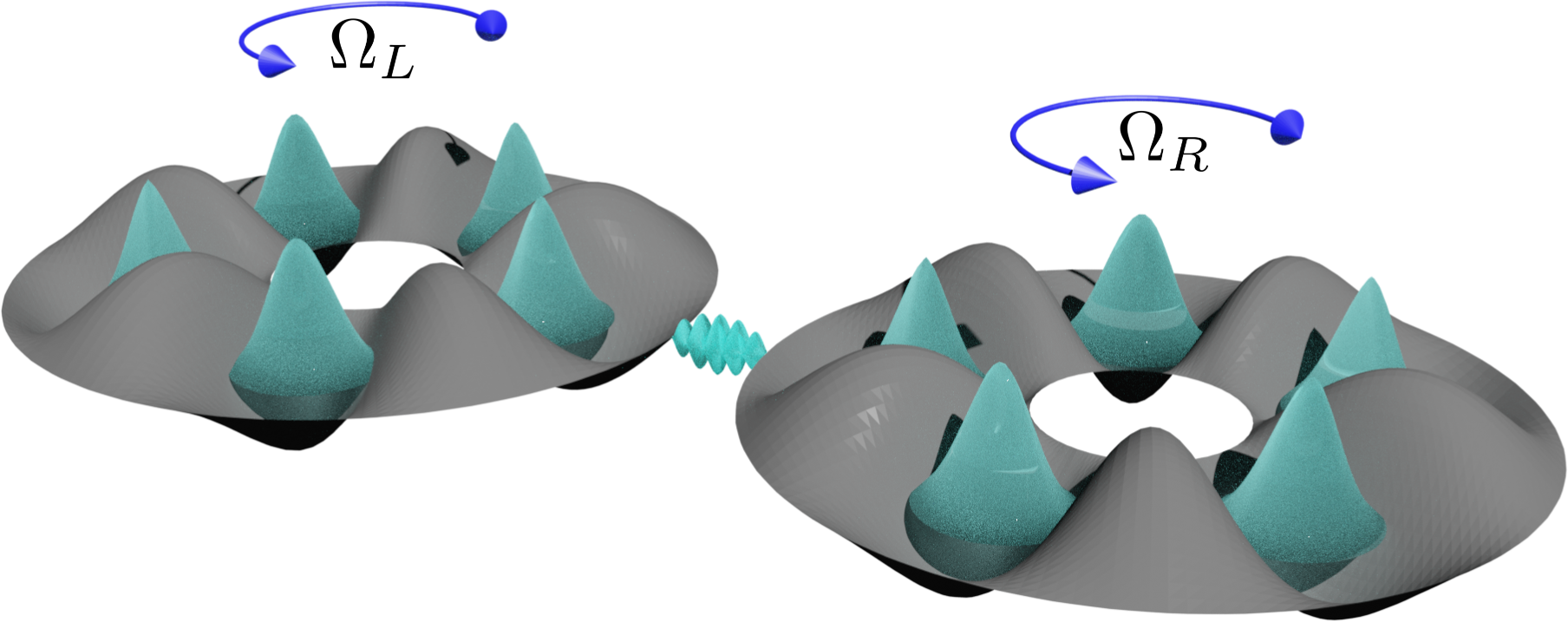}
\caption{Schematic diagram of an optical lattice consisting of two sided rings of five sites each. We represent the small link between rings, $t_l$,  with a small interfering pattern connecting the rings. Both rings have an effective artificial gauge field denoted by $\Omega_L$ and $\Omega_R$.
}
\label{fig:diagram-sided-rings}
\end{figure}

Relying on the considerable know-how achieved in magneto-optic circuit design and atom manipulation techniques,    integrated atomtronic circuits define an interesting direction of the field \cite{amico2021roadmap,amico2021atomtronic}. 
In this context, simple circuits of coupled rings and wave guides have been considered \cite{amico_superfluid_2014,safaei2019monitoring,escriva2021static,richaud2017quantum,aghamalyan2013effective,polo2016geometrically,pelegri2019second,bland2020persistent}. Coupled ring condensates, in particular, are  prototype systems  for the definition of coherent cold atom networks in which  matter wave flows are manipulated as a resource.    Even though several attempts have been done in this direction, the analysis carried out so far show  that independent winding numbers can coexist in the two rings separately \cite{pelegri2019second,bland2020persistent}. The transfer of winding numbers, though, has not been achieved. 

Here, we solve the above bottleneck  and  demonstrate a coherent QPS in two coupled rings arranged side-by-side as in Fig. \ref{fig:diagram-sided-rings}. In the two rings, matter-wave currents with different winding numbers are assumed to be  imparted through effective magnetic flux \cite{dalibard2011colloquium}.   We will argue that to observe such  effect it is important that the system  works  in the full-fledged quantum regime.  Indeed, the QPS we observe results from an oscillation of entangled states of angular momentum states of the two rings.

\paragraph*{The model.}
We consider a system of bosonic atoms trapped in two coupled
coplanar  rings lattice, each  of $N_s$ sites, and subjected to an effective magnetic flux $\Omega_\alpha$, $\alpha=L,R$.
See Fig.~\ref{fig:diagram-sided-rings} for a schematic picture
of the system.
The system's Hamiltonian  reads
\begin{eqnarray}
&&H= H_L+H_R + H_I(\tau) \;,  \nonumber \\
 &&H_\alpha\!=\!\!\!\sum_{i=0}^{N_s-1}\!\!\left[
 U(\hat{n}_{\alpha,i}^2-\hat{n}_{\alpha,i})
 -t
\left(
e^{- i\frac{2\pi\Omega_\alpha}{N_s}}\hat{a}_{\alpha,i}^\dagger\hat{a}_{\alpha,i+1}
+h.c.
\right)\right]\!,
\nonumber\\&&
H_I(\tau)=-t_{l}(\tau)\left(
\hat{a}_{L,0}^\dagger\hat{a}_{R,0}
+\hat{a}_{R,0}^\dagger\hat{a}_{L,0}
\right).
\label{eq:HBH}
\end{eqnarray}
$\hat{a}_{\alpha,i}$ and $\hat{a}_{\alpha,i}^\dagger$ annihilate and create, respectively, a boson in the  site $i$ of the ring $\alpha$, satisfying periodic conditions $\hat{a}_{\alpha,N_s}=\hat{a}_{\alpha,0}$, $\hat{a}_{\alpha,N_s}^\dagger=\hat{a}_{\alpha,0}^\dagger$, and the bosonic number operator is $\hat{n}_{\alpha,i}=\hat{a}_{\alpha,i}^\dagger\hat{a}_{\alpha,i}$.
The parameters $t$ and $U\geq0$ are the intra-ring hopping amplitude and repulsive interaction; $t_l$ describes the tunneling among the two rings.  We set $t=1$.

The limits $U=0$ and $U\to\infty$ can be treated  analytically for $t_l=0$.
The former reduces to the one particle problem, while the latter
maps to tight-binding hard-core bosons\cite{wu2006bose}.
We focus on the regime with weak inter-ring coupling $t_l<t$, allowing us to work out both the numerical simulations obtained by exact diagonalization and  perturbative analysis in $t_l/t$.

For uncoupled symmetric rings, $t_l=\Omega_L=\Omega_R=0$, 
the energy spectrum is invariant under the exchange of left and right states, or under the inversion of angular momentum directions in each separate ring, resulting in specific degeneracies in the  spectrum of (\ref{eq:HBH}).

For $\Omega_L=\Omega_R>0$, 
the rotational symmetry is broken, and clockwise and counterclockwise
current states split into separate energy levels. 
In this regime,  the eigenstates are found either non-degenerate or two-fold degenerate.
A finite $t_l$ splits the remaining degenerate energy levels,
and the new eigenvectors become a superposition of the uncoupled 
degenerate states.
We shall see that such level splitting is  important for the formation of specific entangled states enabling the QPS (see supplemental for an example of such energy splittings).

\begin{figure}[tbh]
  \centering
  \includegraphics[width = \linewidth]{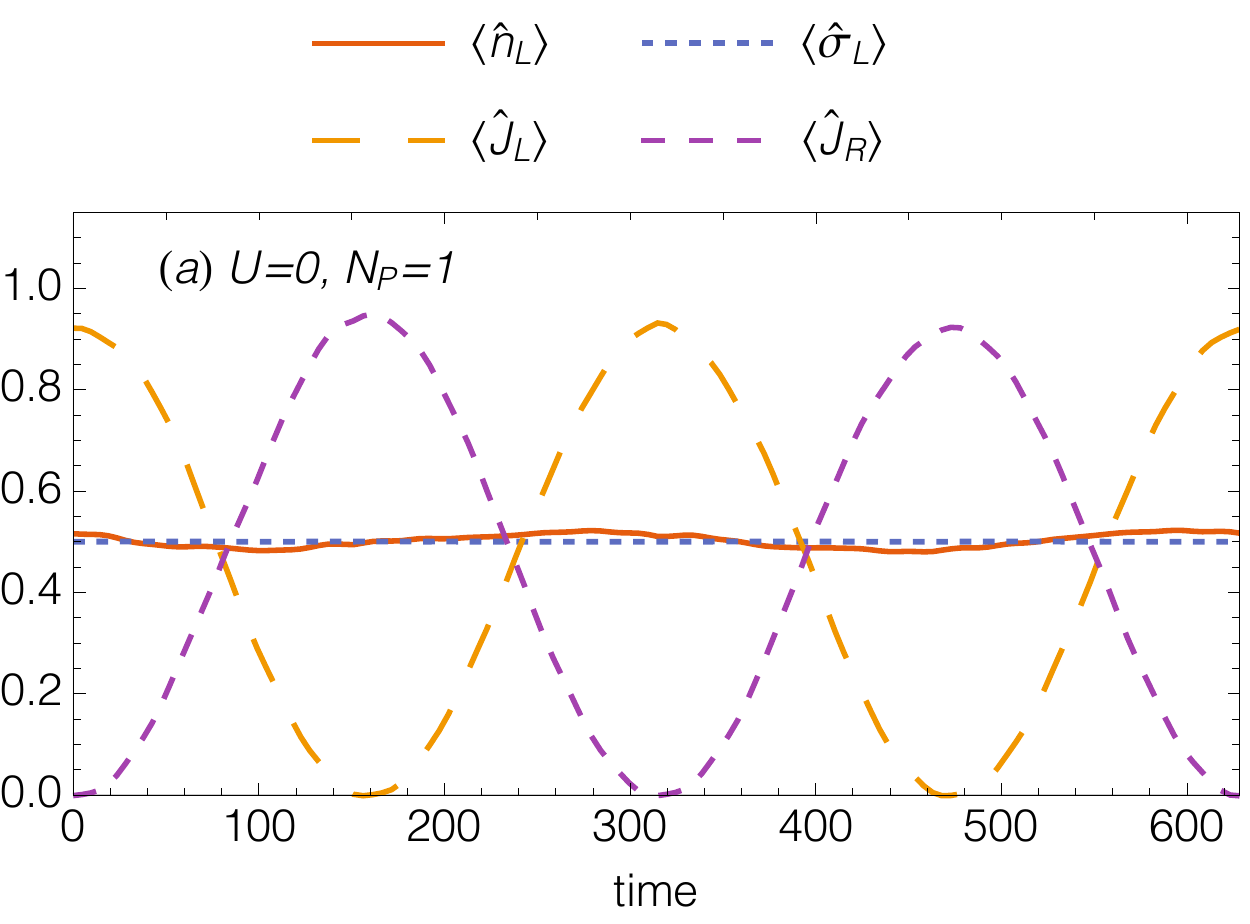}
  \includegraphics[width = \linewidth]{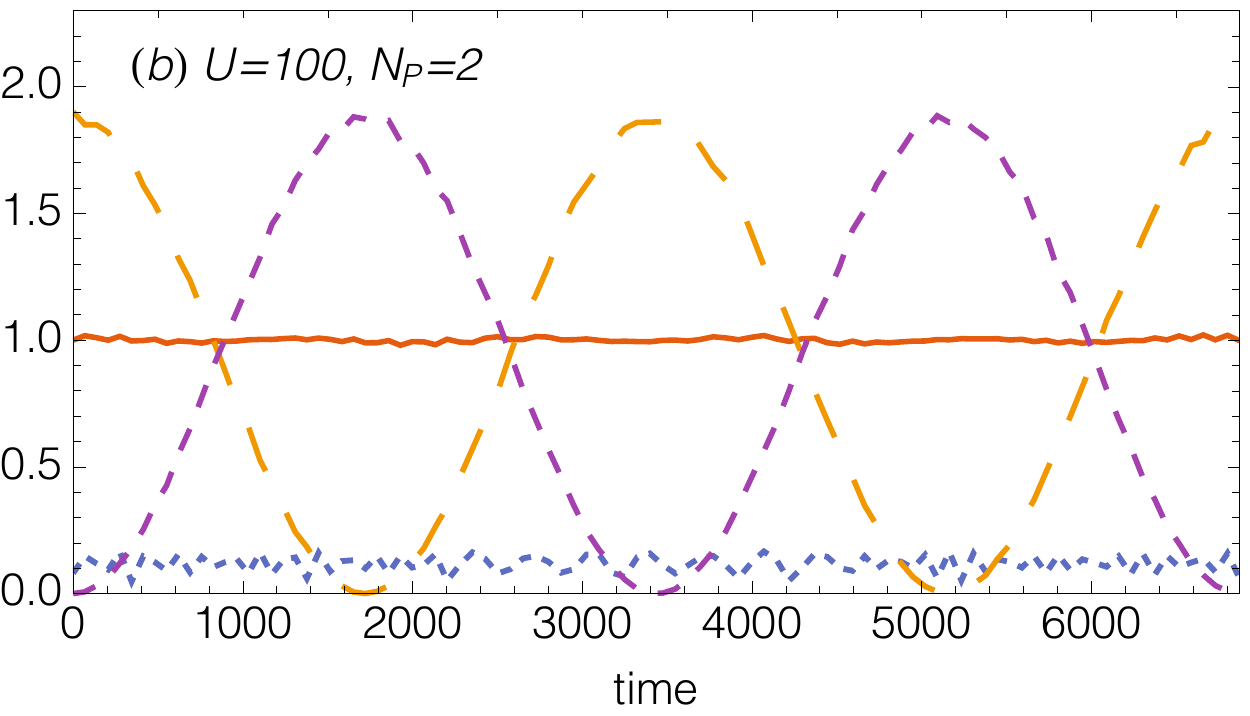}
\caption{Particle number's expectation, standard deviation, and current oscillations 
after quenching a system of two rings of fives sites and
initial fluxes $\Omega_L=1$, $\Omega_R=0$. The final velocities
and interactions, particle number and $t_l$ are
(a) $\Omega_L=\Omega_R=0.05$, $U=0$, $N_p=1$, $t_l=0.05$ and
(b) $\Omega_L=\Omega_R=0.1$,
$U=100$, $N_p=2$, $t_l=0.1$ ($N_p>2$ is considered in Fig.~\ref{fig:particleholesymmtetry}).}
\label{fig:current-oscillations}
\end{figure}

\paragraph*{Quench Protocol.}
We first construct a state 
with the same density of particles in each ring,
$\frac{N_p}{2}$,
but with zero angular momentum on the right, $\langle J_{R}\rangle = 0$, and a non vanishing current 
on the left, $\langle J_{L}\rangle\doteq J_{max}$. We denote such state as  $|L_{j};R_0\rangle\doteq|\frac{N_p}{2},J_{max};\frac{N_p}{2},0\rangle$, resulting to be   an eigenstate of the system when $\Omega_L=j$, $\Omega_R=0$, with $j$ integer.
Next, we perform the quench on the system to 
$\Omega_L=\Omega_R\gtrsim0$.
Then, the  state  $|L_{j};R_0\rangle$
is a superposition of quasi-degenerate eigenvectors,
each with the same particle number $\frac{N_p}{2}$
and total current $J_{max}$.
After the quench, a time evolution occurs on the expectation value of the  number of particles $\langle \hat{n}_\alpha\rangle$,
its variance $\sigma^2_\alpha$, and the current $\langle J_\alpha\rangle$, defined as
\begin{eqnarray}
&&\sigma^2_\alpha=\langle \hat{n}_\alpha^2\rangle-\langle \hat{n}_\alpha\rangle^2\,,  \\
&&\langle J_\alpha\rangle=-i\sum_{k}\langle
\hat{a}_{\alpha,k}^\dagger\hat{a}_{\alpha,k+1}
-\hat{a}_{\alpha,k+1}^\dagger\hat{a}_{\alpha,k}\rangle \,,
\end{eqnarray}
with $\hat{n}_\alpha=\sum_i \hat{n}_{\alpha,i}$.

\begin{figure}[tbh]
  \centering
  \includegraphics[width = \linewidth]{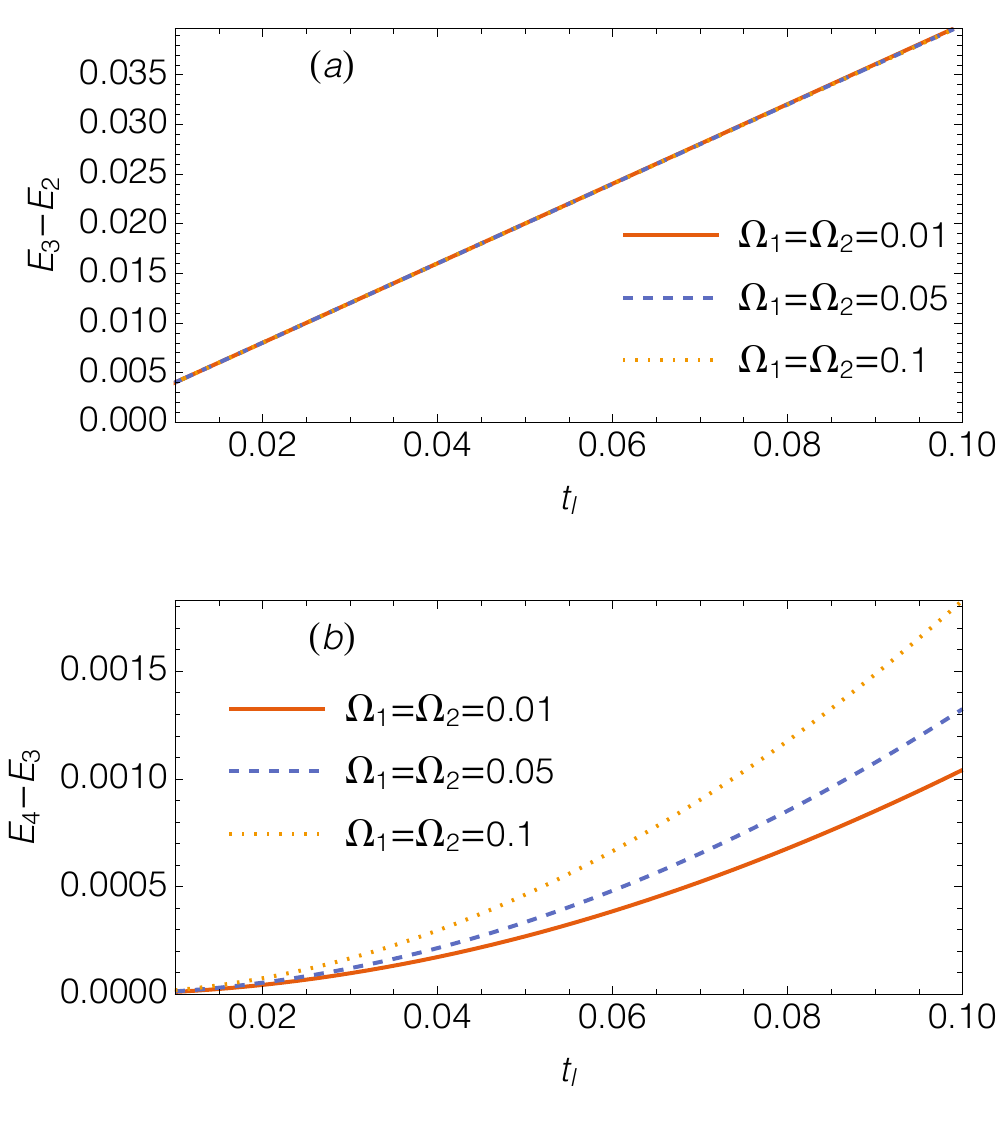}
\caption{Energy gap 
(such that $T=\frac{2\pi}{\Delta E}$, see supplemental material)
as a function of the interring coupling $t_l$
and $\Omega_L=\Omega_R=0.01,0.05,0.1$
for: (a) one particle ($\Delta E=\frac{2 t_l}{N_s}$), (b) two particles and $U=100$
($\Delta E\propto t_l^2$).
}
\label{fig:period-tl}
\end{figure}

\paragraph*{Results.}
Current oscillations between left and right rings
are found for small and large interactions,
$U\lesssim0.01$ and $U\gtrsim10$,
when following the quench.
See Fig.~\ref{fig:current-oscillations} for the cases $U=0$ and $U=100$.
The currents in each ring oscillate completely out of phase,
between a maximum value $J_{max}$ and zero:
\begin{equation}
|L(\tau);R(\tau) \rangle =\cos\left(\omega\,\tau\right)|L_{j};R_0\rangle+
\sin\left(\omega\,\tau\right)|L_0;R_{j}\rangle,
\end{equation}
with $\omega=\frac{\Delta E}{2\pi}$,  $\Delta E$ being the energy gap between the involved states.
Importantly,  
the expectation number  $\langle \hat{n}_\alpha\rangle$ 
in each of the two rings results to be nearly constant at all times:
{\it no net transfer of particles between rings
occur and current oscillations happen due to the phase slipping
through the weak link}. Note that the phase by itself
does not carry any angular momentum or direction, and for the
current to change an arbitrary  small flux
$\Omega_{L/R}=\Omega_\alpha$ is required to be applied in each ring.

\begin{figure}[tbh]
  \centering
    \includegraphics[width = 1 \linewidth]{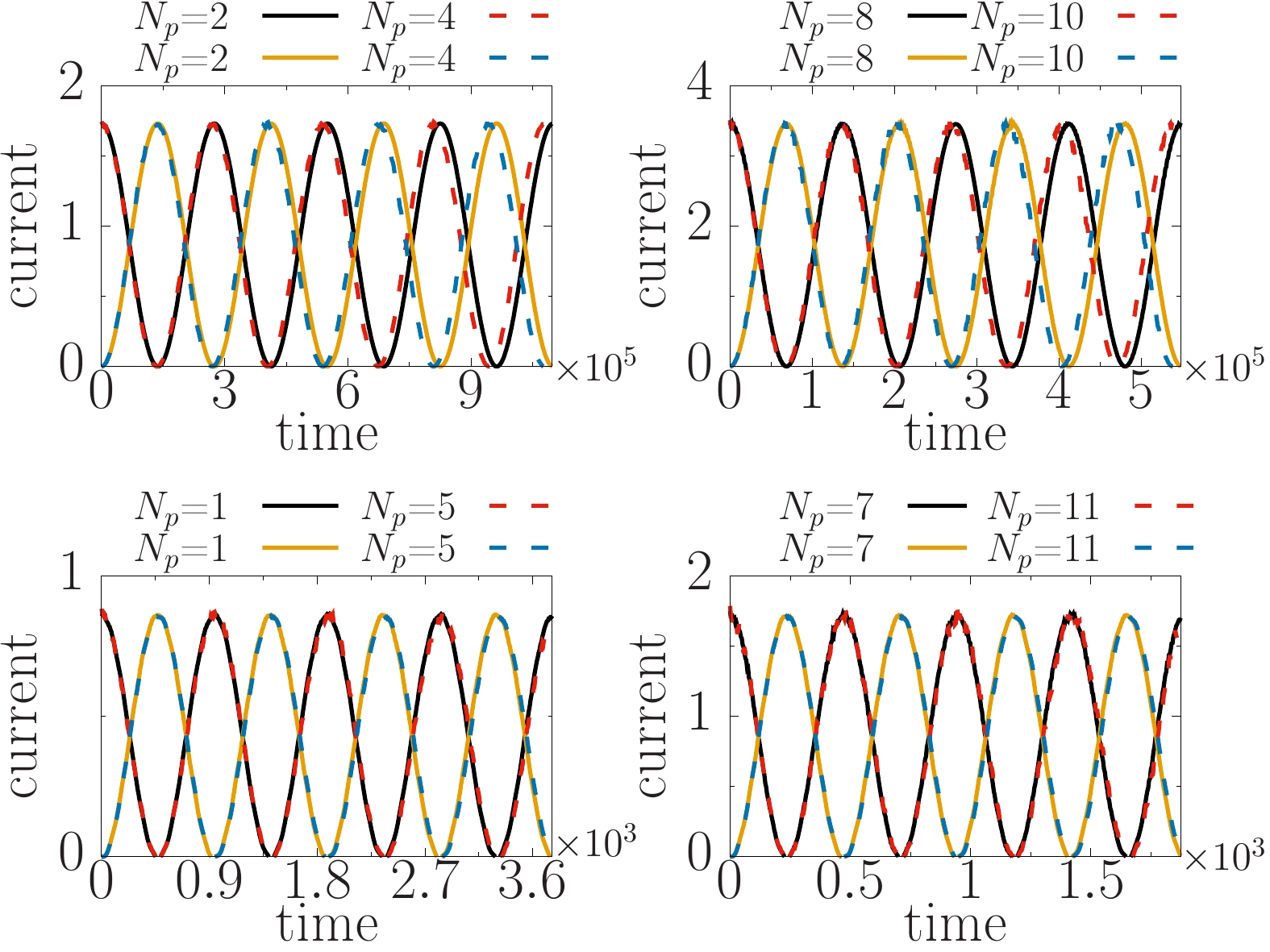}
\caption{Each plot compares QPS where (excess) particles $N_e$ and holes $2N_s-N_e$
are exchanged in the strongly interacting regime, $U=1000$, and in rings of $N_s=3$ sites. 
Particle-hole symmetry is displayed for two different commensurate fillings,
such that the total number of particles is $N_p= 2N_s\times N_b+N_e$, with $N_b=0$
(left plots) and $N_b=1$ (right plots), and $N_e=\{2,4\}$ (top plots) and $N_e=\{1,5\}$
(bottom plots). We quench from $\Omega_L=1$, $\Omega_R=0$, to $\Omega_L=0.01$,  $\Omega_R=0.01$, and from $t_l=10^{-5}$ to $t_l=0.01$. Note that the period substantially changes for each commensurate filling $N_b=0,1$, despite having the same number of particles and holes. These changes appear because of the energy levels shift proportionally to $N_b$.
}
\label{fig:particleholesymmtetry}
\end{figure}

The specific particle configuration
$n_\alpha$ and $\sigma_\alpha^2$, and the maximum current $J_{max}$
in $|L_j;R_0\rangle$ and $|L_0;R_j\rangle$
depend on $U$ and $N_p$ (see Table \ref{tab:summary} and supplemental material). 
By using perturbation analysis, we find that $\omega$ depends linearly in $t_l$ for $U=0$, and quadratically for large $U$
(except for odd $N_p$, see Fig.~\ref{fig:period-tl} and Table~\ref{tab:summary}).
Important insights on the effect of the interaction can be obtained by studying the limit $U\to\infty$.
Although the expected number of particles results to be barely affected by $U$, its variance $\sigma_\alpha$ does. For $U=0$,  $\sigma_\alpha^2=\frac{N_p}{4}$. For large interactions, instead, any measurement
of the occupation would always find half the particles in each ring, and therefore  $\sigma_\alpha^2=0$ is found
(or $\frac{N_p\pm1}{2}$ particles in each ring and $\sigma_\alpha^2=\frac14$ for $N_p$ odd).
We also remark that, because of the  particle-hole symmetry holding for large interactions, QPS
for  holes occur similarly to the particles ones (see Fig.~\ref{fig:particleholesymmtetry} and supplemental material). As for the maximum current, in the case $U=0$,  it results to scale linearly with the number of particles,
$J_{max}=2\,N_p\sin\left(\frac{2\pi}{N_s}\right)$,
while for large interactions we find
$J_{max}=4\cos\left(\frac{\pi}{N_s}\right)
\sin\left(\frac{\pi N_p}{N_s}\right)$.
To transfer larger currents, one can simply start with an integer flux $\Omega_L$ in $1<n<k_\textrm{max}$. 
Numerical tests for various ring sizes and initial currents corroborate 
that QPS are still found as $N_p$ and angular momenta are increased (see Fig.~\ref{fig:particleholesymmtetry} and supplemental material).

\begin{table}[tbh] 
\begin{center} 
\begin{tabular}{cc|c|c|c|c} 
  & &$\langle \hat{n}\rangle$  &  $\sigma^2$  & $J_{max}$  & $T$ 
\\\hline 
$U=0$  && $\frac{N_p}{2}$   &   $\frac{N_p}{4}$   &  $2\,N_p\sin\left(\frac{2\pi}{N_s}\right)$  & $\frac{\pi N_s}{t_l}$
\\\hline
\multirow{2}{*}{$U\to\infty$} & $N_p$ even & \multirow{2}{*}{$\frac{N_p}{2}$}   &   $0$   &  \multirow{2}{*}{$4\cos\left(\frac{\pi}{N_s}\right)
\sin\left(\frac{\pi N_p}{N_s}\right)$}  & $\propto\frac{1}{t_l^2}$
\\
  & $N_p$ odd &   &   $\frac{1}{4}$   &    & $\propto\frac{1}{t_l}$
\end{tabular}  
\caption{
Expected value of the occupation and variance in each ring, amplitude and period 
of current oscillations, in the case of no interactions and for $U\to\infty$.
} 
\label{tab:summary} 
\end{center} 
\end{table} 

By controlling $t_l$ in time, and relying on our condition of weak ring-ring coupling, we note that different entangled states of angular momenta can be engineered by our scheme.
The transfer of angular momentum can be obtained by  
manipulating $t_l$ on $\tau=\frac{T}{2}\times(2n+1)$.
For  $\tau=\frac{T}{4}\times (2n+1)$, for example,  the entangled state 
$|L_j;R_0\rangle+|L_0;R_j\rangle$
can  be achieved.

\begin{figure}[htb!]
  \centering
    \includegraphics[width = 1 \linewidth]{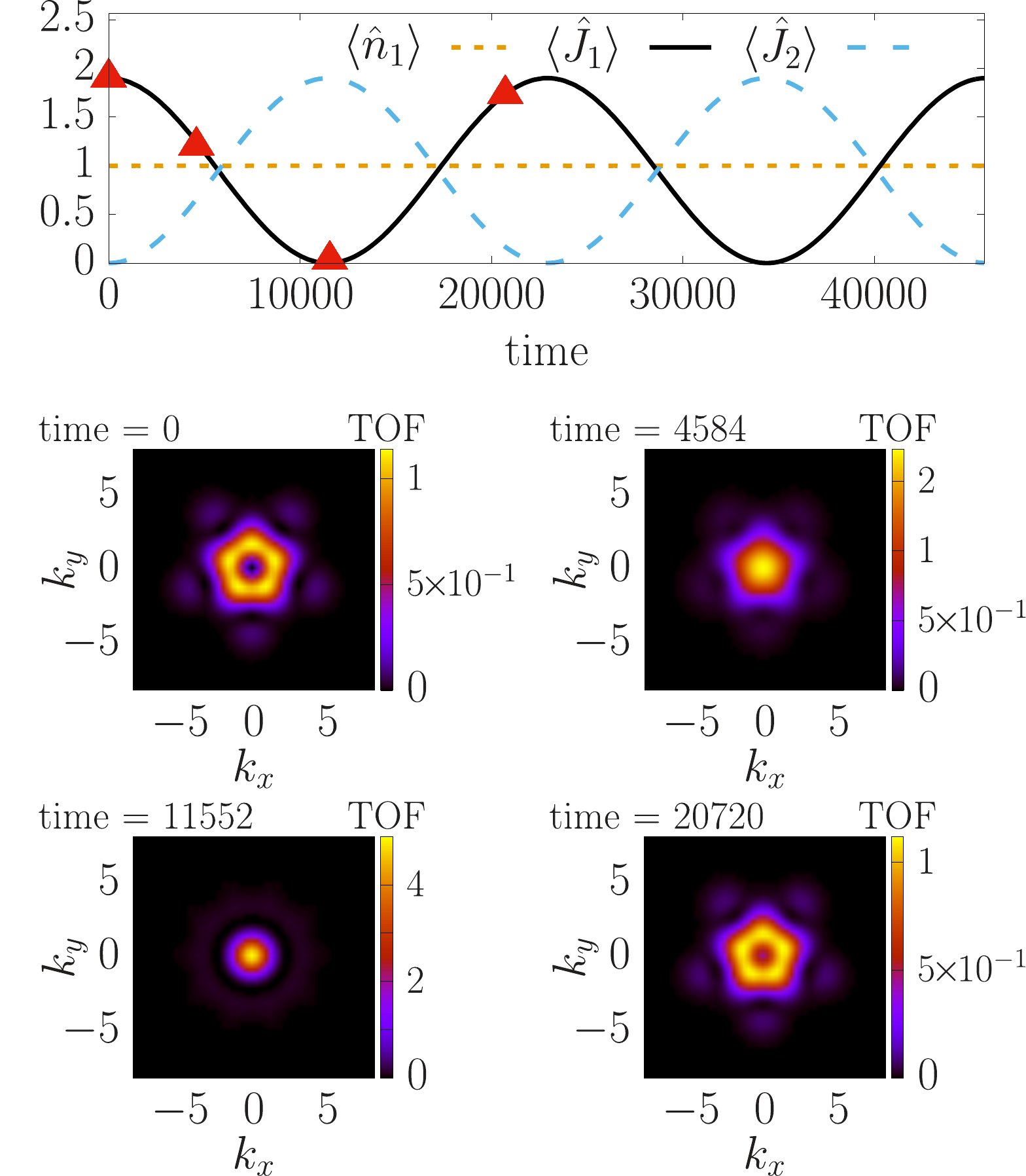}
\caption{
Time-of-flight expansion of the left ring. Parameters are: $\Omega_L=1,\;\Omega_R=0$ to $\Omega_L=0.01,\;\Omega_R=0.01$, $t_l=0$ to $t_l=0.05$, $N_p=2$, $N_s = 5$ in each ring, $U=10$. Triangles indicate the times at which the TOF snapshots where taken. 
}
\label{fig:TOF}
\end{figure}

\paragraph*{Read-out of the QPS.}
Matter wave currents can be detected through  time-of-flight (TOF) measurements \cite{moulder2012quantized,amico_quantum_2005}. Such measurements in the far field are directly related to the  momentum distribution at the moment in which the condensate is released from the ring trap. Therefore, the time evolution of the persistent current is reflected in the time evolution of the momentum distribution: $n({\bf k},t) = \sum_{i,j} \: e^{i\bf k\cdot ({\bf R}_i-{\bf R}_j)}  C_{i,j}(t) $ with $C_{i,j}(t) = \langle \psi(t) | a^{\dagger}_ia_j | \psi(t) \rangle$ being the one-body correlation function between different sites and $\textbf{R}_{j}$ denoting the position of the lattice sites of the ring.
In each ring, we find that  $n({\bf k},t)$ evolves from a peak momentum distribution to the characteristic circular-shaped one as soon as the system acquires one unit of angular momentum. Such dynamics in the TOF provides the read-out of the transfer of  coherent phase slips between zero and one unit of angular momentum. See  Fig.~\ref{fig:TOF}.

\paragraph*{Discussion and Conclusions.}
We have theoretically demonstrated QPS between two tunnel coupled rings of interacting bosons: We prepare two different phase-states of the two separated rings; after quenching the tunnel between the rings, we observe a coherent oscillation between the phase states with nearly vanishing population fluctuations (in each of the two rings). Once calibrated, the scheme can be used to produce, transfer, and entangle current states by tuning on and off the the weak link at specific times after the quenching protocol. We find that the phase slips faster from one ring to the other for stronger inter-ring couplings.
Interactions reduce the maximum current in each ring, and make phase slips slower (see supplemental material for the interplay between interactions and period of oscillations). Indeed, such phenomenon occurs as a direct  consequence of the entangled state created between the phase states of the two rings (macroscopic superposition of all particles rotating with different angular momenta in each ring). The coherent oscillations of the QPS are characterized by the simultaneous creation and destruction of current states in each ring (see Sec. A of supplemental material for specific examples). As such, QPS transfer is a genuine quantum effect that cannot be captured by standard mean-field analysis such as Gross-Pitaevskii based approach. In fact, the latter neglects entanglement and, to the best of our knowledge, cannot describe  coherent transfer of matter-wave without transfer of population (which is an essential trait of our demonstration).  We note  that the coupling between the rings is perturbative and as such, the corresponding emergence of quasi-degenerate states involving a superposition of left and right current states hold for large particle numbers. The scope of our results can be further enlarged by resorting to a suitable particle-hole symmetry. Therefore, our QPS  are expected to occur also in systems with large particle numbers.
We studied the momentum distribution that is the standard method to analyze neutral matter-wave currents in cold atoms experiments \cite{moulder2012quantized,amico_quantum_2005,greiner2002quantum,gerbier2005interference,kato2008sharp,Hoffmann2009visibility}

Our work provides a specific  platform to observe QPS  in cold atoms that is a well known open problem in the field. At the same time, our work results are relevant to progress in the implementation of integrated atomtronic circuits \cite{amico2021roadmap,amico2021atomtronic}. Specifically, our results effectively enable  atomtronic circuit based on coupled atomic rings: In a sense analogue to the ‘Rapid Single Flux Quantum Logic’ conceived with SQUID’s \cite{likharev1991rsfq},  complex structures where the information is encoded in the phase slips inherent to the different rings could be implemented.

{\it Acknowledgements}. We thank Gianluigi Catelani and Wayne Jordan Chetcuti for discussion. A. P-O. acknowledges financial support from Secretaria d’Universitats i Recerca del Departament d'Empresa i Coneixement de la Generalitat de Catalunya
cofunded by the European Union Regional Development Fund
within the ERDF Operational Program of Catalunya
(project QuantumCat, ref. 001-P-001644).

\bibliographystyle{apsrev4-1}

\bibliography{references.bib}

\clearpage

\appendix

\section{Single particle analytical results in Fourier space}
\label{app:fourier}

For one particle and $\Omega_L=1$, the initial state is
$|\psi_+\rangle=\frac{1}{\sqrt2}(b_{L,1}^\dagger+ b_{R,0}^\dagger)|0\rangle$,
where $b_{\alpha,l}^\dagger=\frac{1}{\sqrt{N_s}}\sum_j e^{i\frac{2\pi (j\,l)}{N_s}}a_{\alpha,j}^\dagger$ creates a particle with momentum $l$ in the ring $\alpha$.
It has an average particle number per ring $n_\alpha=\frac12$ and variance $\sigma_\alpha^2=\frac14$.
In this case, $H_{BH}$ in momentum space consists of
a diagonal term,
where the eigenvectors are vortex states, $b_{\alpha,i}^\dagger|0\rangle$,
 and a perturbative one,
 that transfers momentum modes between left and right rings,
\begin{align}
 H_{BH}=&
 -2\,\sum_{\alpha=L,R}\sum_{i=0}^{N_s-1}
\cos\left(2\pi (i-\Omega_\alpha)/N_s\right)\hat{b}_{\alpha,i}^\dagger\hat{b}_{\alpha,i}
\nonumber\\&-
 \frac{t_l}{N_s}\sum_{i,j=0}^{N_s-1}
\left(\hat{b}_{L,i}^\dagger\hat{b}_{R,j}+h.c.\right).
\label{eq:HBH2}
\end{align}
The energy gap is $\Delta E=\langle\psi_-|H_p|\psi_-\rangle-\langle\psi_+|H_p|\psi_+\rangle=\frac{2t_l}{N_s}+\mathcal{O}(t_l^2)$,
where $H_p$ is the second term of Eq.~(\ref{eq:HBH2})
and $|\psi_-\rangle=\frac{1}{\sqrt2}(b_{L,1}^\dagger- b_{R,0}^\dagger)|0\rangle$.
If $U=0$, each particle behaves independently, and current
oscillations are due to the simultaneous but opposite transfer
of the ground and first excited momentum modes.
 The particle distribution is binomial,
with $n_\alpha=\frac{N_p}{2}$ and variance $\sigma_\alpha^2=\frac{N_p}{4}$.
The initial (and maximum) current scales with the number of particles,
and evaluates to $J_{max}=2\,N_p\sin\left(\frac{2\pi}{N_s}\right)$.

\section{Two-state level simplification}
\label{app:two-level}

All the current oscillations found can be understood in terms of dynamics within
a system of two quasi-degenerate states.
For a given set of fluxes, $\Omega_L=\Omega_R=\Omega\gtrsim0$, and a weak enough
interring coupling, $t_l\gtrsim0$, pairs of quasi-degenerate states
can effectively be treated as uncoupled from the rest of eigenstates.
Here we show this explicitly with two specific examples, one for $U=0$,
which reduces to the case of one particle, and another for large interactions,
$U\to \infty$.

\begin{figure}[tb]
  \centering
    \includegraphics[width = 0.95 \linewidth]{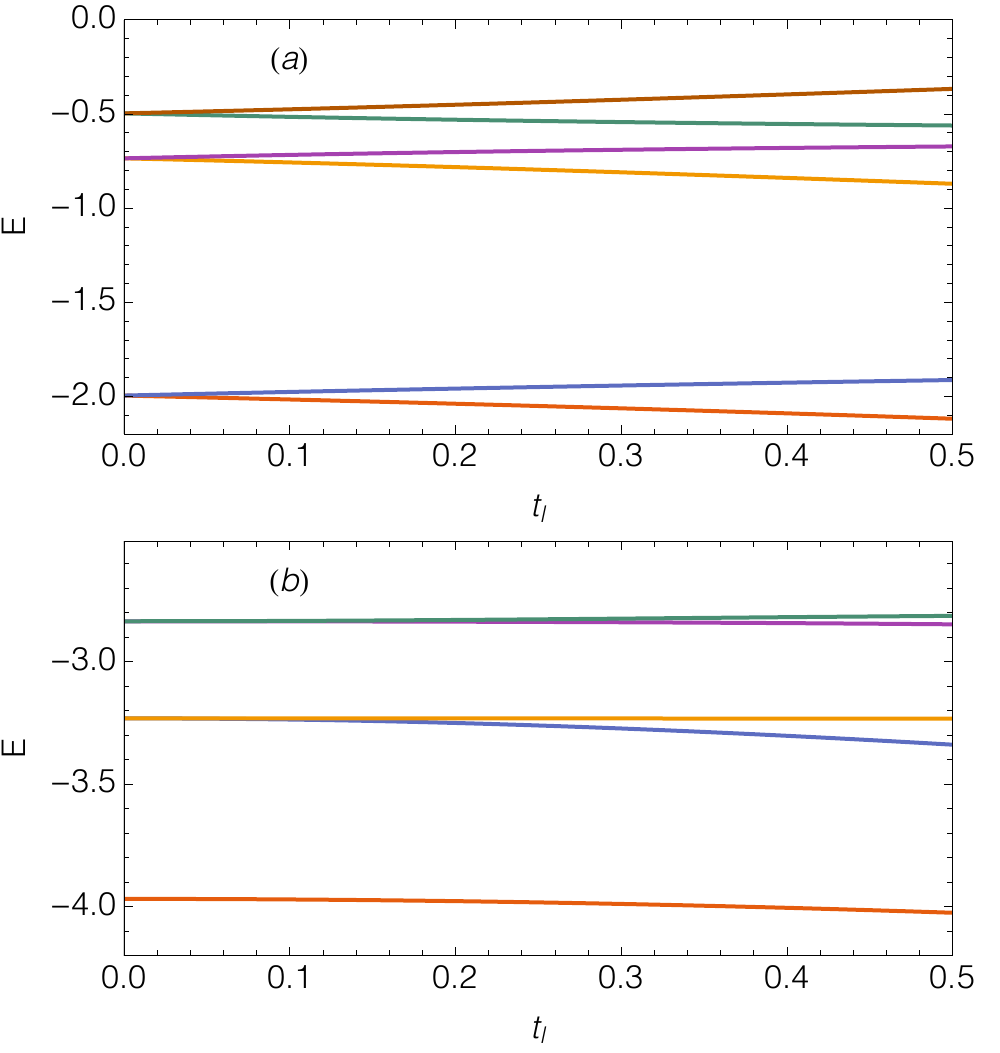}
\caption{Energy levels as a function of the interring coupling $t_l$
for (a) $U=0$, $\Omega_1=\Omega_2=0.05$, $N_p=1$ and (b) $U=100$, $\Omega_1=\Omega_2=0.1$, $N_p=2$ .
In (a), the corresponding eigenstates are, from bottom to top,
$|\phi_0^+\rangle$, $|\phi_0^-\rangle$, $|\phi_1^+\rangle$, $|\phi_1^-\rangle$,
$|\phi_{-1}^+\rangle$, and $|\phi_{-1}^-\rangle$, as described in the main text.
The first and second energy splittings are separated by
$E_1-E_0=-2\cos\left(2\pi (1-\Omega)/N_s\right)+2\cos\left(2\pi (\Omega)/N_s\right)$, while the gap between the second and third
is $\Delta\epsilon_\Omega=
\frac{8\pi}{N_s}\sin\left(\frac{2\pi\,j}{N_s}\right)\Omega+\mathcal{O}(\Omega^3)$.
In (b), the energy levels correspond to eigenstates
$|\chi_0\rangle$, $|\chi_0^+\rangle$, $|\chi_0^-\rangle$,
$|\chi_{1}^+\rangle$, and $|\chi_{1}^-\rangle$.
}
\label{fig:spectra-1p-2p}
\end{figure}

In the case of one particle,
terms coupling different momentum modes in Eq.~(\ref{eq:HBH2}) enter
at order $\mathcal{O}(t_l)$ when diagonalizing, assuming single ring energy levels
$E_{i}=-2\,t \cos\left(2\pi (i-\Omega)/N_s\right)$
are separate enough. Neglecting those terms, $H_{BH}$ can be written in terms
of two-mode independent Hamiltonians,
\begin{align}
H_{BH}^{(i)}=&
E_{i}\left(\hat{b}_{L,i}^\dagger\hat{b}_{L,i}+\hat{b}_{R,i}^\dagger\hat{b}_{R,i}\right)
\nonumber\\&
- \frac{t_l}{N_s}\left(\hat{b}_{L,i}^\dagger\hat{b}_{R,i}+h.c.\right).
\end{align}
Within this picture, the ground and first excited states are
$|\phi_0^{\pm}\rangle=\frac{1}{\sqrt{2}}|0,1\rangle\pm\frac{1}{\sqrt{2}}|1,0\rangle$, where $|N_L,N_R\rangle$ define states with $N_L$ static particles
in the left ring and $N_R$ in the right one.
Equivalently, the next excited states are
$|\phi_1^{\pm}\rangle=\frac{1}{\sqrt{2}}|0,1^+\rangle\pm\frac{1}{\sqrt{2}}|1^+,0\rangle$.
Here the subindex $1$ in $\phi$ and superindex $+$ in $1$
indicate one unit of angular momentum in the direction of $\Omega$.
These pair of eigenstates are energetically separate from 
$|\phi_0^{\pm}\rangle$, and from the next excited states,
$|\phi_{-1}^{\pm}\rangle=\frac{1}{\sqrt{2}}|0,1^-\rangle\pm\frac{1}{\sqrt{2}}|1^-,0\rangle$,
in which the rotation is opposite to the fluxes $\Omega$ (see Fig.~\ref{fig:spectra-1p-2p}~(a)).
These energy gaps are much larger than the energy separating $|\phi_i^-\rangle$
and $|\phi_i^+\rangle$ states in each pair, which is $\Delta E=\frac{2t_l}{N_s}$
in all cases.
The evolution of the initial state from our protocol,
$|\psi_+\rangle=\frac{1}{\sqrt{2}}|1^+,0\rangle+\frac{1}{\sqrt{2}}|0,1\rangle$,
consists of two independent evolutions with the same frequency
$\omega=\frac{\Delta E}{2\pi}$, one for $|1^+,0\rangle$,
and another for $|0,1\rangle$,
\begin{align}
|\psi(\tau)\rangle&=
\frac{e^{-i\,E_1\,\tau}}{\sqrt{2}}\left[
\cos\left(\omega\,\tau\right)|1^+,0\rangle
+i\sin\left(\omega\,\tau\right)|0,1^+\rangle
\right]
\nonumber\\&+
\frac{e^{-i\,E_0\,\tau}}{\sqrt{2}}\left[
\cos\left(\omega\,\tau\right)|0,1\rangle
+i\sin\left(\omega\,\tau\right)|1,0\rangle
\right]
\nonumber\\&\doteq
\cos\left(\omega\,\tau\right)|L_{1};R_0\rangle+
\sin\left(\omega\,\tau\right)|L_0;R_{1}\rangle.
\end{align}

For large interactions and even $N_p$,
eigenstates have the same number of particles $N=\frac{N_p}{2}$
in each ring.
The ground state is $|\chi_0\rangle=|N,N\rangle$, energetically separate from
the the first excited states, among which
are $|\chi_1^{\pm}\rangle=\frac{1}{\sqrt{2}}|N,N^+\rangle\pm\frac{1}{\sqrt{2}}|N^+,N\rangle$, and
$|\chi_{-1}^{\pm}\rangle=\frac{1}{\sqrt{2}}|N,N^-\rangle\pm\frac{1}{\sqrt{2}}|N^-,N\rangle$. In this case the energy gap $\Delta E$ separating $|\chi_{i}^{+}\rangle$ 
and $|\chi_{i}^{-}\rangle$ states is
much smaller and proportional to $t_l^2$, see Fig.~\ref{fig:spectra-1p-2p}
for the case with $N=1$ in each ring.
Given an initial state $|\psi(t=0)\rangle=|N^+,N\rangle$, it evolves 
with a relatively low frequency $\omega=\frac{\Delta E}{2\pi}$,
\begin{align}
|\psi(\tau)\rangle&=
e^{-i\,E_1\,\tau}\left[
\cos\left(\omega\,\tau\right)|N^+,N\rangle
+i\sin\left(\omega\,\tau\right)|N,N^+\rangle
\right]
\nonumber\\&\doteq
\cos\left(\omega\,\tau\right)|L_{1};R_0\rangle+
\sin\left(\omega\,\tau\right)|L_0;R_{1}\rangle.
\end{align}

\section{Spectrum for large interactions}
\label{app:spectrum}

The spectrum for $U=0$ does not depend on the number of particles, $N_p$, and the relevant energy gap and evolution after the quench can be computed analytically with first order perturbation theory.
In contrast, for large interactions, the energy gaps do depend
on $N_p$, and scale as $t_l^2$, instead of $t_l$, which implies that second-order perturbation theory is needed. In this section we analyze how the spectrum for large interactions depends on $\Omega_\alpha$, $t_l$, $U$, and $N_p$.

\begin{figure}[tb]
  \centering
  \includegraphics[width = \linewidth]{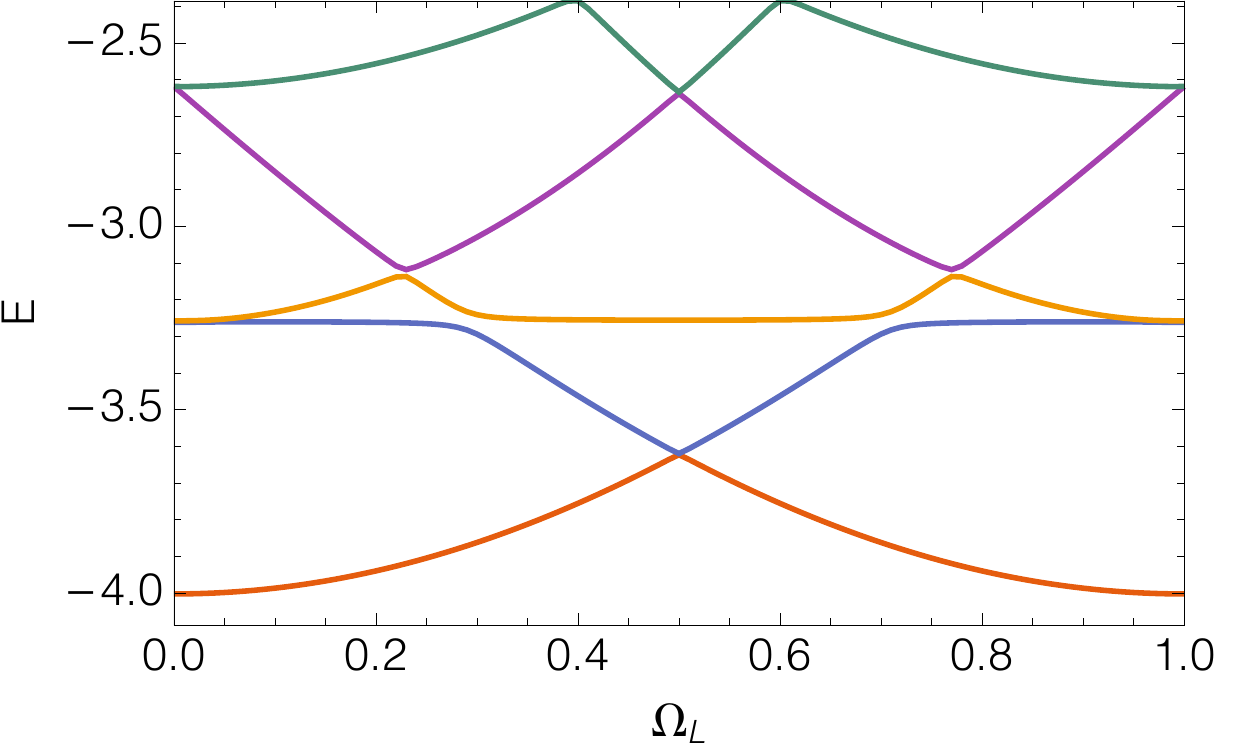}
\caption{Spectrum of energies depending on $\Omega_L$ (left) for $U=100$, $N_p=2$, $t_l=0.1$, and $\Omega_R=0$.
A quench can be made such that the ground state at $\Omega_L=1$
is a superposition of the third and fourth excited states 
at $\Omega_L=0$.
}
\label{fig:spectrum-omega}
\end{figure}

The energy spectrum for two particles and $\Omega_R\gtrsim0$ as a function of $\Omega_L$ is plotted in Fig.~\ref{fig:spectrum-omega}.
At $\Omega_L=\Omega_R\gtrsim0$, states are symmetric with respect to
the exchange of left and right rings. They correspond to, from bottom to top
and in the same notation as in Appendix~\ref{app:two-level},
\begin{align}
|\chi_0\rangle=&|1,1\rangle, & &
\nonumber\\
|\chi_0^+\rangle=&\frac{|0,2\rangle+|2,0\rangle}{\sqrt2},
&
|\chi_{0}^-\rangle=&\frac{|0,2\rangle-|2,0\rangle}{\sqrt2},
\nonumber\\
|\chi_1^+\rangle=&\frac{|1,1^+\rangle+|1^+,1\rangle}{\sqrt2},
&
|\chi_1^-\rangle=&\frac{|1,1^+\rangle-|1^+,1\rangle}{\sqrt2},
\nonumber\\
|\chi_{-1}^+\rangle=&\frac{|1,1^-\rangle+|1^-,1\rangle}{\sqrt2},
&
|\chi_{-1}^-\rangle=&\frac{|1,1^-\rangle-|1^-,1\rangle}{\sqrt2}.
\end{align}
As $\Omega_L$ increases, this symmetry is broken, and eigenvectors turn to, approximately,
$|1,1\rangle$,
$|0,2\rangle$, $|2,0\rangle$,
$|1^+,1\rangle$, $|1,1^+\rangle$,
$|1,1^-\rangle$, $|1^-,1\rangle$.
The energies increase quadratically with $\Omega_L$,
except for the state in which the left ring is empty, $|0,2\rangle$, in which
the energy remains constant.
At $\Omega_L\lesssim1$, we have the same states
except for a general shift in one unit of angular momentum
in the right ring. As $\Omega_L$ decreases from $\Omega_L=1$, states with particles
on the left also increase their energy quadratically.
Due to the small but finite $t_l$, the energy levels have avoided crossings
(some are too small to see in the figure). In terms of the relation
between the ground state at $\Omega_L=1$, $|1^+,1\rangle$, the initial state in our protocol,
with the states at $\Omega_L\gtrsim0$, these avoided crossings imply
$|1^+,1\rangle=\frac{1}{\sqrt{2}}|\chi_1^+\rangle+\frac{1}{\sqrt{2}}|\chi_1^-\rangle$.

\begin{figure}[tb]
  \centering
  \includegraphics[width = \linewidth]{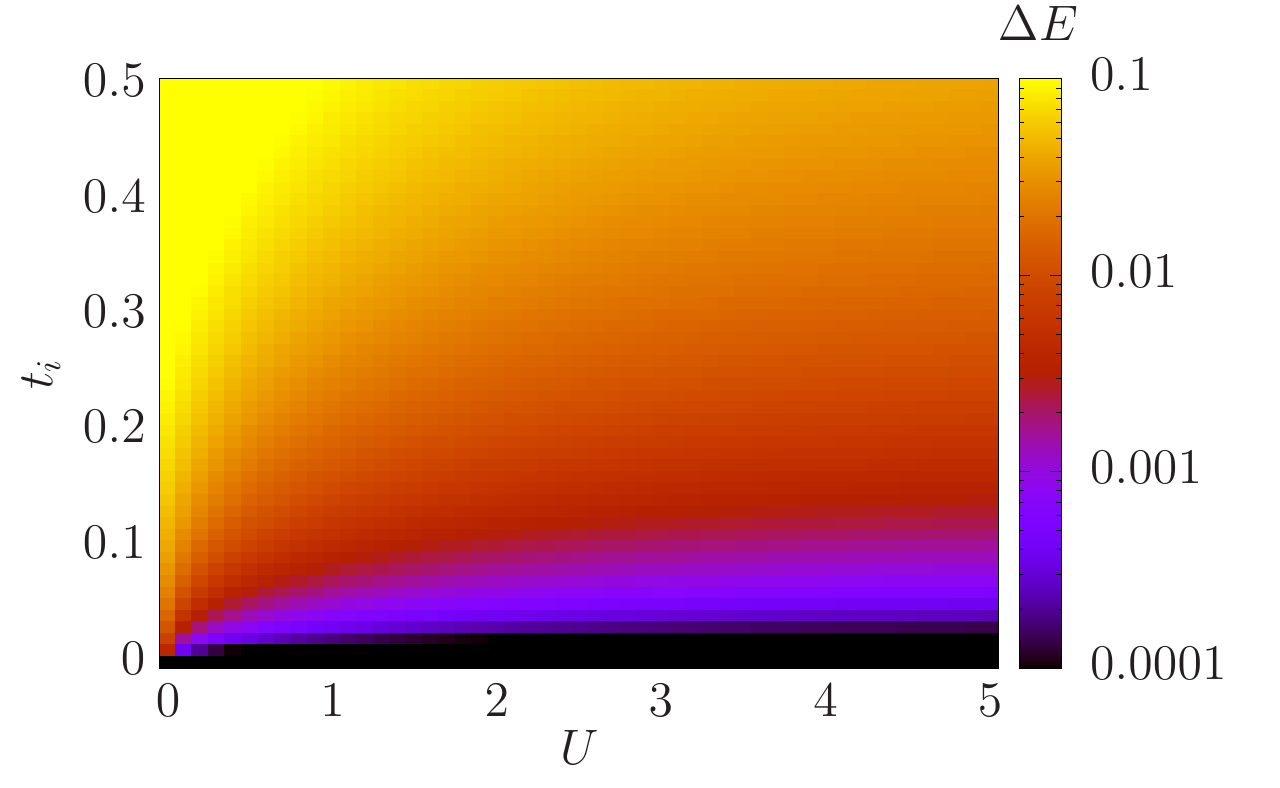}
\caption{
Energy gap of the two occupied states as a function of the link strength $t_i$ and interactions $U$ after a quench starting from $\Omega_L=1$, $\Omega_R=0$ to $\Omega_L=\Omega_R=\Omega_j=0.1$
for $N_p=2$ particles and $N_s = 5$.
All curves collapse for large enough interactions and the gap behaves equally as a function of $t_l$. }
\label{fig:gap-tl}
\end{figure}

The corresponding energy gap, $\Delta E=E_4-E_3$, is plotted as a function
of $U$ and $t_l$ in Fig.~\ref{fig:gap-tl}. 
At $U=0$, $\Delta E$ is relatively large and shows no dependence on $t_l$. In this case, the energy gap between the third and fourth excited states corresponds to $\Delta E_\Omega$ (See Fig.~\ref{fig:spectra-1p-2p}).
As interactions $U$ increase, the energy gap $\Delta E$
closes and its dependence on $t_l$ converges to a fixed gradient.

As for the dependence of $\Delta E$ on $N_p$, it exhibits two main features.
First, the energy gaps for $N_p$ even are much smaller than for $N_p$ odd,
the former being proportional to $t_l^2$ and the latter to $t_l$.
Second, the energy gaps are the same under the exchange of particles and holes.
For large interactions, the BH model maps to a spinless
FH model, in which the spectrum is invariant, except for a general shift, under the exchange $N_p \leftrightarrow 2N_s-N_p$.
This implies that, in the case of rings with five sites,
the spectra and energy gaps are the same for the pairs of particle numbers $N_p=1$, 9; 2, 8; 3, 7; and 4, 6.
Apart from the symmetry under the exchange $N_p \leftrightarrow 2N_s-N_p$, we find the energy gap is largest at smallest or largest occupations,
and decreases down to a minimum at half filling.

Particle-hole symmetry still holds when $N_p>2N_s$, as shown in
\fref{fig:particleholesymmtetry} in the main text.
If there are more particles than sites, such that there
is a number $2N_s\times N_b$ of particles evenly distributed in the background, and $N_e$ excess particles,
$N_p=2N_s\times N_b+N_e$, the energy gap as a function of $N_e$ has the same qualitative behavior.
Therefore, ring systems with more particles than sites have very similar behavior in terms of phase slips.
If $N_e=0$, as in the case of full filling, the ground state
has $N_b$ particles per site, and is energetically very separate from the first excited state (the gap being proportional to $U$, which we assume large). In this case, 
our protocol does not allow for phase slips.

\clearpage
\end{document}